\documentclass[aps,pre,groupedaddress,showpacs,preprint]{revtex4}
\usepackage{graphicx}
\usepackage[dvips]{epsfig}
\usepackage{dcolumn}
\usepackage{subfigure}%
\usepackage{bm}
\usepackage{amssymb}
\usepackage{amsmath}

\begin{document}
\title{Stochastic resonance due to internal noise in reaction kinetics}

\author{Carlos Escudero}

\affiliation{Mathematical Institute, University of Oxford, 24-29 St Giles', Oxford OX1 3LB, United Kingdom}

\begin{abstract}
We study a reaction model that presents stochastic resonance purely due to internal noise. This means that the only source of fluctuations comes from the discrete character of the reactants, and no more noises enter into the system. Our analysis reveals that the phenomenon is highly complex, and that is generated by the interplay of different stochasticity at the three fixed points of a bistable system.
\end{abstract}

\pacs{64.60.My, 05.40.-a, 05.45.-a, 82.20.-w}
\maketitle

Stochastic resonance is a phenomenon that has been extensively studied in the literature, both for its inherent interest, as well as for its broad range of applications~\cite{jung}. Since the seminal works on this subject appeared~\cite{benzi}, much theoretical work has been done in order to extend the variety of situations in which similar phenomena arise. Very interesting examples, as stochastic coherence~\cite{gang}, the effect of coloured noise~\cite{hanggi}, multiplicative noise~\cite{fulinski}, and noise-mediated localization~\cite{cabrera} have already been found and analyzed. A different and very relevant question is how to get some type of stochastic resonance purely due to internal fluctuations. One answer was given in the form of system size resonance~\cite{casa}, obtained when the optimal output of a system is achieved for a certain finite number of constituent subsystems. Stochastic optimization also appears by tuning a continuous parameter: numerical studies have shown that chemical reactions can undergo stochastic coherence due to internal noise, both in autonomous~\cite{jung2} and non-autonomous situations~\cite{reichl}. But however it is remarkable the seemingly little attention that has been paid on internal noise amplification of an external signal in chemical kinetics mimicking the classical situation in stochastic resonance, apart from the seminal work by Dykman {\it et al.}~\cite{dykman}. The main goal of the present work is to study this phenomenon by means of analytical tools, and without relying on the assumption of Gaussian noise, what allows us to study the problem even if one state is almost empty.

Reaction kinetics is a prototypical problem where the effect of the fluctuations can be quantified~\cite{gardiner}. Internal fluctuations in these systems appear due to the discrete character of the reactants, and a mean-field description always omits some of their features. We will show that stochastic resonance is one of them. For our purposes we will consider the following series of reactions: $\emptyset \to X$, at rate $K_1$, $X \to \emptyset$, at rate $K_2$, $2X \to 3X$, at rate $K_3$, and $3X \to 2X$, at rate $K_4$. We can interpret this problem as a population dynamics model: the first reaction represents emigration into the system, the second death, the third reproduction, and the last competition for limited resources of nutrient. A similar problem but without the three body reaction (the competition reaction) was studied in Ref.~\cite{elgart}, and we will show here that including this new reaction opens the possibility of stochastic resonance.
This system can be modelled by means of a combinatorial master equation~\cite{gardiner}
\begin{eqnarray}
\nonumber
\frac{\partial P(n,t)}{\partial t}=K_1 [P(n-1,t)-P(n,t)]+K_2 [(n+1)P(n+1,t)-nP(n,t)]+ \\
\nonumber
\frac{K_3}{2}[(n-1)(n-2)P(n-1,t)-n(n-1)P(n,t)]+ \\
\frac{K_4}{6}[(n+1)n(n-1)P(n+1)-n(n-1)(n-2)P(n,t)].
\end{eqnarray}
To study this master equation we will employ the techniques developed by Elgart and Kamenev~\cite{elgart}. Consider the generating function
\begin{equation}
G(p,t)=\sum_{n=0}^\infty p^n P(n,t),
\end{equation}
this function obeys the imaginary time Schr\"{o}dinger equation $\partial_t G=-\hat{H}G$, where the Hamiltonian is given by
\begin{equation}
\label{hamilton}
\hat{H}(\hat{p},\hat{q})=K_1(1-\hat{p})+K_2(\hat{p}-1)\hat{q}+\frac{K_3}{2}(1-\hat{p})\hat{p}^2\hat{q}^2+\frac{K_4}{6}(\hat{p}-1)
\hat{p}^2\hat{q}^3.
\end{equation}
The effect of the momentum and coordinate operators is $\hat{p}G(p,t)=pG(p,t)$ and $\hat{q}G(p,t)=-\partial_p G(p,t)$. The detailed procedure is described in~\cite{elgart}, but it is worth pointing out that similar techniques were previously introduced in the literature~\cite{doi}. We can write now the classical equation of motion, that is given by
\begin{equation}
\label{meanf}
\dot{q} \equiv \frac{dq}{dt}=\left. -\frac{\partial H}{\partial p} \right|_{p=1}=K_1-K_2q+\frac{K_3}{2}q^2-\frac{K_4}{6}q^3,
\end{equation}
in which the coordinate $q$ plays the role of mean-field density. In order to understand the phenomenon of stochastic resonance in a chemical system we need some kind of bistability. For this, we make the next assumptions on the rate constants: $K_1=\epsilon L n^3$, $K_2= n^2(\epsilon+L+\epsilon L)$, $K_3=2n(1+\epsilon+L)$, and $K_4=6$, where $0<\epsilon<1$, $n>0$, and $L>1$. This way Eq. (\ref{meanf}) has three fixed points: $q_{-}=\epsilon n$, $q_0=n$, and
$q_+=L n$, $q_{-}$ and $q_+$ are stable and $q_0$ is unstable. These new parameters $\epsilon$ and $L$ are auxiliary mathematical variables that are introduced in order to simplify the notation; they have, in contrast, no direct physical meaning but the one given through their relation with the rate constants. The variable $n$ is a measure of the size of the system, and will be the tuning parameter that will allow us to find the resonance, because we can modify its value without changing the relative distance between the fixed points. On the other hand, assuming bistability is quite natural in population dynamical systems, as for instance in those modelling epidemics~\cite{ludwig}.

The deterministic equation predicts an evolution to one of the stable fixed points, $q_{-}$ for initial conditions $q(0)<q_0$, and $q_+$ if $q(0)>q_0$. In the stochastic case, both states become metastable, and the system will jump from one to the other indefinitely. In this case the system does not experience the same noise strength in the neighbourhood of the fixed points, since the noise has not been imposed externally in an
{\it ad hoc} manner. To estimate the frequency of jump we will employ the instanton technique for reaction kinetics, developed in the paper by Elgart and Kamenev~\cite{elgart}. Let us briefly explain why we do think that this method is the most appropriate for the present case. It is known that this type of theories, as the one given by
Hamiltonian~(\ref{hamilton}), can be mapped onto a stochastic differential equation~\cite{escudero}; so one could be suggested to derive the stochastic equation and then apply standard techniques~\cite{jung}. However, the presence of cubic powers of the momentum in the Hamiltonian denote the presence of non-Gaussian noise, and its exact correspondence with Langevin equations is still unknown~\cite{lopez}.

The action corresponding to Hamiltonian Eq. (\ref{hamilton}) is
\begin{equation}
S[p,q]=Et-\int_0^t q\dot{p}dt-q(0)[p(0)-1],
\end{equation}
where we have used the fact that $E=H(p,q)$ is an integral of motion, i. e. $\dot{E}=0$. To simplify our calculations we will perform the change of variables $q=\tilde{q}n$ and $t=\tilde{t}/n^2$, so the action becomes
\begin{equation}
S[p,\tilde{q}]=n\tilde{H}(p,\tilde{q})\tilde{t}-n\int_0^{\tilde{t}} \tilde{q}\dot{p}d\tilde{t}-n\tilde{q}(0)[p(0)-1],
\end{equation}
where $\tilde{H}$ denotes the Hamiltonian with the new rate constants, and the fixed points have now the values
$\tilde{q}_-=\epsilon$, $\tilde{q}_+=L$, and $\tilde{q}_0=1$. The tildes will be suppressed from now on.

\begin{figure}
\includegraphics[width=12cm]{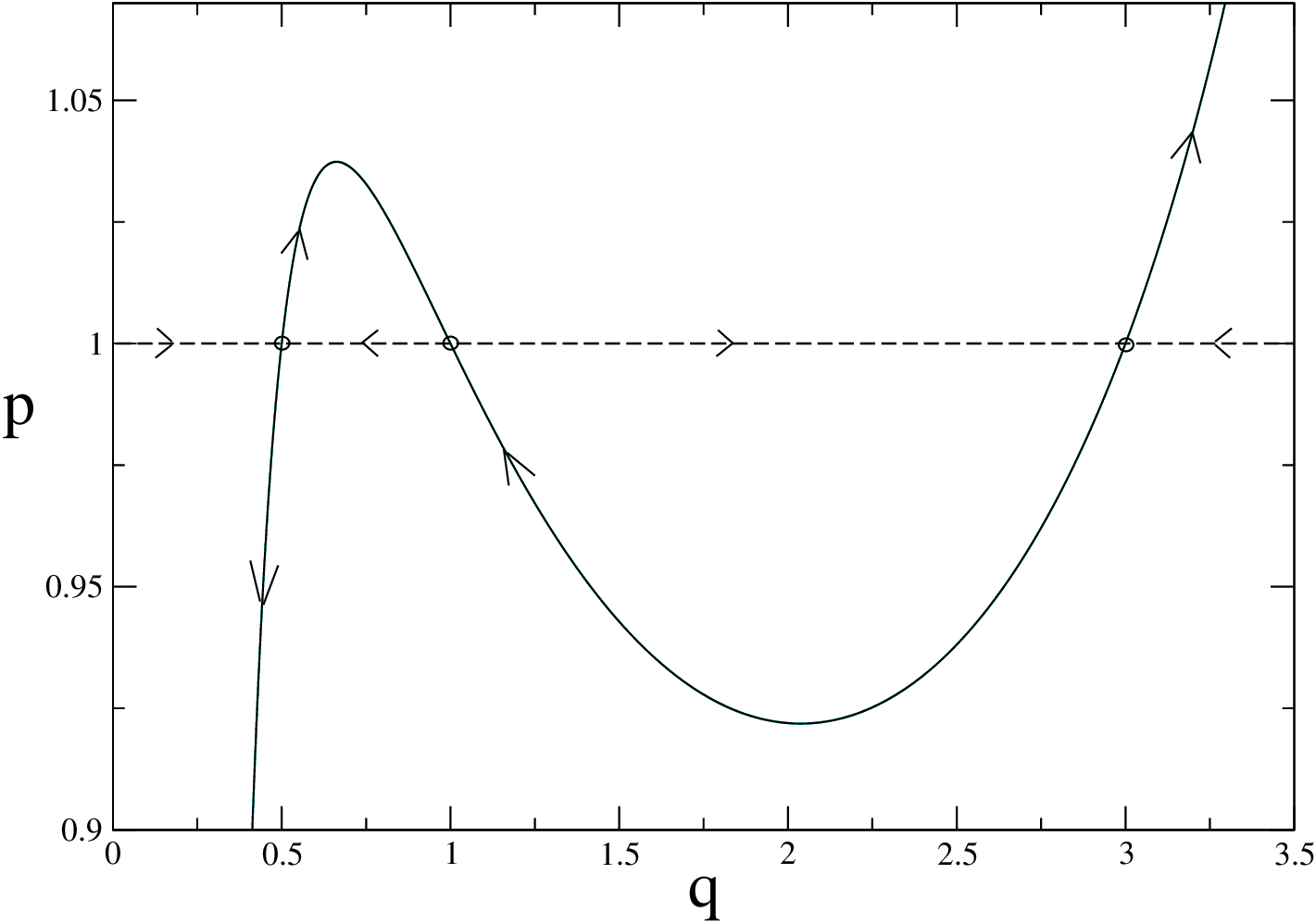}
\caption{Phase space for $\epsilon=1/2$, $n=1$, and $L=3$. The mean-field line is shown dashed and the non-mean-field line is represented as a solid line. The fixed points are encircled and the arrows show the direction of motion along the different lines. Note that if we only consider the mean-field line, then $q=1/2$ and $q=3$ are stable and $q=1$ is unstable. However, taking into account the whole phase space the three fixed points become saddles.}
\label{lines}
\end{figure}

In the long time limit, $t \to \infty$, the system approaches the trajectories of zero energy, $H(p,q)=0$. The decay time from one state to the other is estimated as $\tau=\sigma \exp(S_0)$, where $\sigma$ is the relaxation time to the arrival state and $S_0$ is the action along the non-mean-field line~\cite{elgart}. Let us briefly comment on this point. The equation $H(p,q)=0$ has two solutions, $p=1$, the mean-field line, and the non-mean-field or instanton line, both are depicted in Fig.~\ref{lines} for $\epsilon=1/2$, $n=1$, and $L=3$. One can see that if we only consider the mean-field line, then $q_-$ and $q_+$ are stable and $q_0$ is unstable. But if we consider the whole phase space, then we realize that the three points are actually saddles. The time it takes to go from one point to the other is proportional to the exponential of the action along the line, which is identically zero for mean-field lines. Since we are performing a semiclassical calculation, it is valid only for calculating transition times along non-mean-field lines if $S_0>1$. We will estimate the transition time from $q_-$ to
$q_+$ (and {\it vice versa}) as the time it takes to move along the non-mean-field line, since this time is much longer than the time it takes to perform the mean-field part of the trajectory.

In transitions from $q_{-}$ to $q_0$ the action is
$S_0=n(q_{-}-q_0)+n\int_{q_{-}}^{q_0} pdq$,
where $p$ can be obtained from the relation $H(p,q)=0$:
\begin{equation}
\label{instanton}
p=\left(\frac{(\epsilon+L+\epsilon L)q-\epsilon L}{(1+\epsilon+L-q)q^2}\right)^{1/2}.
\end{equation}
Although this integral can be computed straightforwardly, the resulting expression is cumbersome, and would not help us to understand the phenomenon. So we will study the simpler case $\epsilon \to 0$:
\begin{equation}
\label{s1}
\lim_{\epsilon \to 0}S_0=n(q_{-}-q_0)+2n\sqrt{L} \cdot \mathrm{arccsc}\left(\sqrt{1+L}\right)=2n\sqrt{L} \cdot \mathrm{arccsc}\left(\sqrt{1+L}\right)-n.
\end{equation}
The reader might wonder how the system can leave the empty state since there is no emigration. This is true, because in an empty system there are no fluctuations, and once there, we will stay in it forever. However, the empty state might be metastable (it is very easy to find a set of reactions that rends the empty state unstable), and a perturbation could lead us out of it. So in this case $\epsilon$ should be considered very small ($\epsilon << 1$), but not identically zero. Now it appears clear another advantage of the Elgart-Kamenev approach: it allows us to estimate the effect of internal fluctuations in a state with almost no population, contrary to perturbative approaches~\cite{gardiner}. A simpler expression can be obtained by taking the limit $L \to \infty$: $S_0=n(q_{-}-q_0)+2n=n$. Note that in this case the action reduces to the distance between the fixed points, suggesting that this distance is the key
parameter that rules the transitions between the metastable states. We will see, however, that the dynamics is not as simple as this.

We can also compute the action along the non-mean-field line from $q_+$ to $q_0$
\begin{equation}
\label{s2}
S_0=n(q_{+}-q_0)+n\int_{q_{+}}^{q_0} pdq \to n(L-1)+2n\sqrt{L}\left[ \mathrm{arctan}\left( \sqrt{L} \right) -\mathrm{arccot}\left( \sqrt{L} \right)\right], \qquad \epsilon \to 0.
\end{equation}
In the limit $L \to \infty$ the second term in the right hand side goes like $\sim \sqrt{L}$, so it is irrelevant compared to the first term, that is the difference between $q_0$ and $q_+$. So we have arrived at the same conclusion as in the last paragraph.

Now, if we want to spend the same time in both trips $q_-\to q_0$ and $q_+ \to q_0$ we need to identify expressions
Eq.(\ref{s1}) and Eq.(\ref{s2}), so we are led to solve the transcendental equation
$\sqrt{L}/2+\mathrm{arccot}\left(\sqrt{L}\right)=\mathrm{arccsc}\left(\sqrt{L+1}\right)+
\mathrm{arctan}\left( \sqrt{L} \right)$.
This equation can be solved numerically to get $L^* \approx 5.43$. At this point one is tempted to use the simplified expressions obtained in the limit $L \to \infty$ in order to get a closed expression for $L^*$. However, equalling the most relevant terms in this limit, one arrives at the contradiction $L^*=1$. This means that we have to take into account that the three points are at a finite distance from each other, because the transition times depend on the relative position of the three fixed points altogether. Note that assuming the equality of both decay times is not essential for the system to undergo resonant behaviour, but it simplifies the forthcoming calculations.

\begin{figure}
\includegraphics[width=12cm]{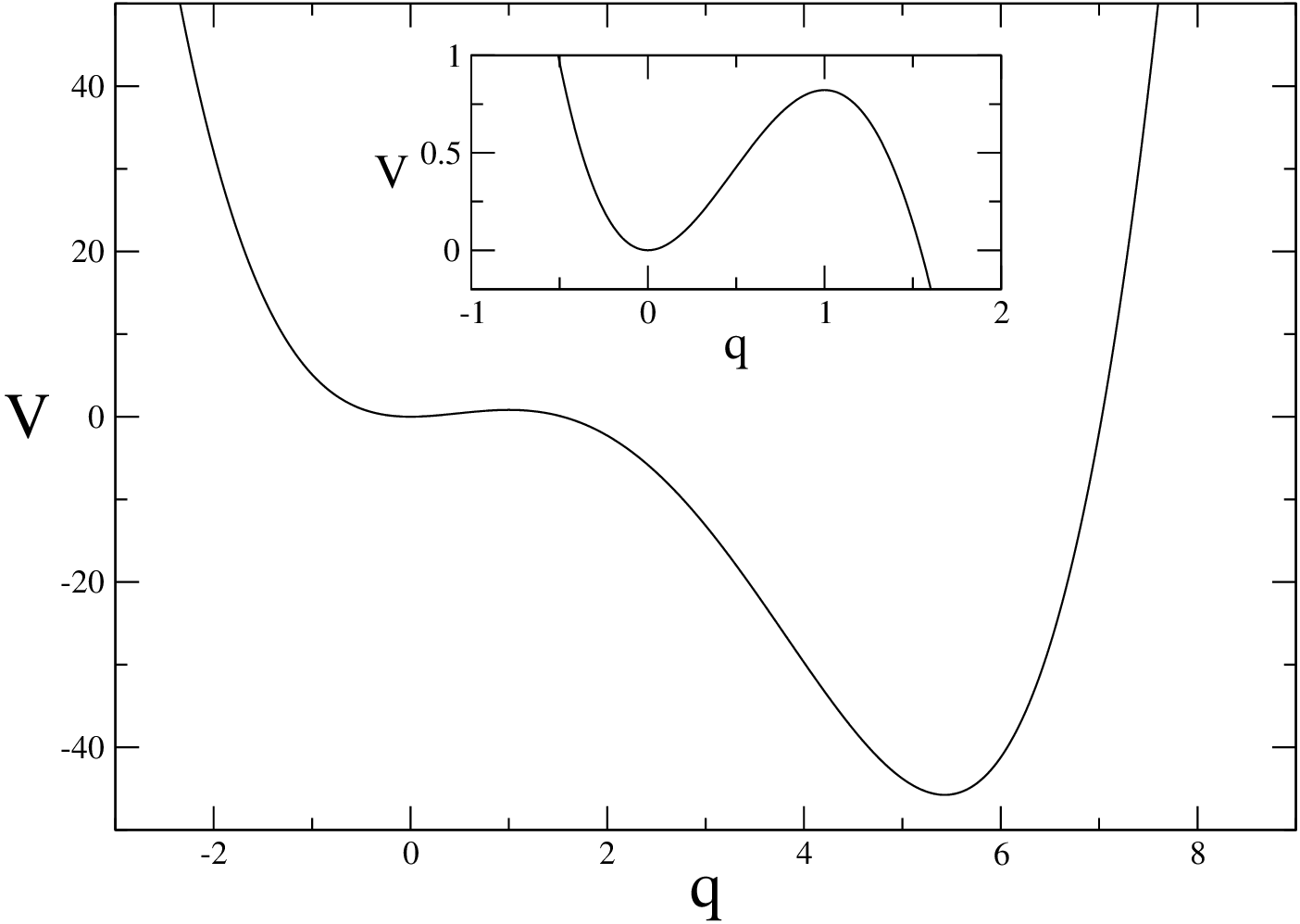}
\caption{Potential energy in the system with $\epsilon=0$, $n=1$, and $L=L^*$ as explained in the text. The minima are displayed at $q=0$ and $q=L^*$, and the maximum is located at $q=1$. One can see the huge difference in depth between the
two wells. The inset shows a closer look to the minimum located at $q=0$, to provide a better idea of what the energy barrier is in this case.}
\label{potential}
\end{figure}

So we finally have our bistable system $(q_-,q_0,q_+)=(0,1,L^*)$. This system is extremely asymmetric, not only because the distance between the fixed points is not same, but more importantly because the "potential energy" at these points is very different. We have defined the potential energy as the integral of the force $V(q)=-\int F(q)dq$, where the force is given by $F(q)=-\partial_q H(p,q)|_{p=1}$. We have depicted the potential in Fig.~\ref{potential}, where one can see the huge difference of depth between the two wells. From this figure one realizes that the fluctuations are much stronger in the
case of a higher population, and so, they have to be stopped by a higher energy barrier. The fluctuations are so weak in the case of a low population, that we have to make it easier for them to cross the barrier. The effect of a nonvanishing third moment in the noise term has been previously examined in the case of a deterministic chaotic "noise"~\cite{chew}. But in this case the weakness of the asymmetry allowed to treat it as a perturbation.

The relaxation time to $q_0$ is given by $\sigma=(L^*-1)^{-1}$, so we have all the ingredients to calculate the decay time
$\tau$, and we can now examine if our system will be resonant to the external signal $f(t)=A\cos\left( \omega t + \phi \right)$. Here $\phi$ is an arbitrary phase and $A$ is the amplitude of the signal. In our population dynamics model it is introduced as a periodic modulation of the competition $K_4 \to K_4+A\cos\left( \omega t + \phi \right)$, and we will assume that the amplitude $A$ of this signal is very small so we can consider it as a perturbation. This perturbation implies
an additional time-dependent term in the Hamiltonian $H_p=n^3 A \cos(\omega n^{-2} t + \phi)(p-1)p^2q^3$,
in non-dimensional variables. The corresponding action reads
\begin{equation}
S_p= n A \cos(\omega n^{-2} t_0) \int_0^t \cos(\omega n^{-2} t)(p-1)p^2q^3 dt
- n A \sin(\omega n^{-2} t_0) \int_0^t \sin(\omega n^{-2} t)(p-1)p^2q^3 dt,
\end{equation}
where we have expressed the phase as a function of the signal initial condition $\phi=\omega n^{-2} t_0$. Now, after assuming a quasistationary variation of the external signal (adiabatic approximation), we can reduce the action to
\begin{equation}
\label{spertur}
S_p= n A \cos(\omega n^{-2} t) \int_0^t (p-1)p^2q^3 dt,
\end{equation}
where we have modified the notation $t_0 \to t$. To compute action (\ref{spertur}) we need to solve the classical equations of motion, $\dot{q}=-\partial_p H$ and $\dot{p}=\partial_q H$,
in the instanton line Eq.~(\ref{instanton}); for the first equation we get
$2dt=\left(L^*q-\sqrt{L^*q^3(1+L^*-q)}\right)^{-1}dq$.
Plugging this last expression together with Eq.~(\ref{instanton}) (note that now $\epsilon=0$ and $L=L^*$) into Eq.~(\ref{spertur}) we obtain, for transitions from $q=0$ to $q=1$
\begin{eqnarray}
\nonumber
S_p= n A \cos(\omega n^{-2} t) \int_0^1 \frac{\sqrt{L^*q}}{2(1+L^*-q)^{3/2}} dq = \\
n A \cos(\omega n^{-2} t) \left( 1-\sqrt{L^*} \cdot \mathrm{arccsc}\left(\sqrt{L^*+1}\right) \right) = n A_1 \cos(\omega n^{-2} t),
\end{eqnarray}
and for transitions from $q=L^*$ to $q=1$
\begin{eqnarray}
\nonumber
S_p= n A \cos(\omega n^{-2} t) \int_{L^*}^1 \frac{\sqrt{L^*q}}{2(1+L^*-q)^{3/2}} dq= \\
n A \cos(\omega n^{-2} t) \left( 1-L^*+\sqrt{L^*}\left[\mathrm{arctan}\left( \sqrt{L^*} \right)-\mathrm{arccot} \left( \sqrt{L^*} \right)\right] \right)= -n A_2 \cos(\omega n^{-2} t),
\end{eqnarray}
where the effective amplitudes $A_1$ and $A_2$ were defined in order to preserve their positivity. The ratio between both actions $\left| S_p(L^* \to 1)/S_p(0 \to 1) \right| \approx 48$ shows that the effective amplitude of the external signal is very different at the two stable fixed points, being much stronger in $q=L^*$ than in $q=0$. It seems that this mechanism is reminiscent of that of noise amplification, that is responsible in turn of the asymmetry of the potential in Fig.~(\ref{potential}). So we can finally write the time-dependent escape rates
\begin{subequations}
\begin{eqnarray}
\tau_1(0 \to 1)= \exp[S_0+S_p(0 \to 1)] \approx \tau(1+n A_1 \cos(\omega n^{-2} t)), \\
\tau_2(L^* \to 1)= \exp[S_0+S_p(L^* \to 1)] \approx \tau(1-n A_2 \cos(\omega n^{-2} t)).
\end{eqnarray}
\end{subequations}

\begin{figure}
\includegraphics[width=12cm]{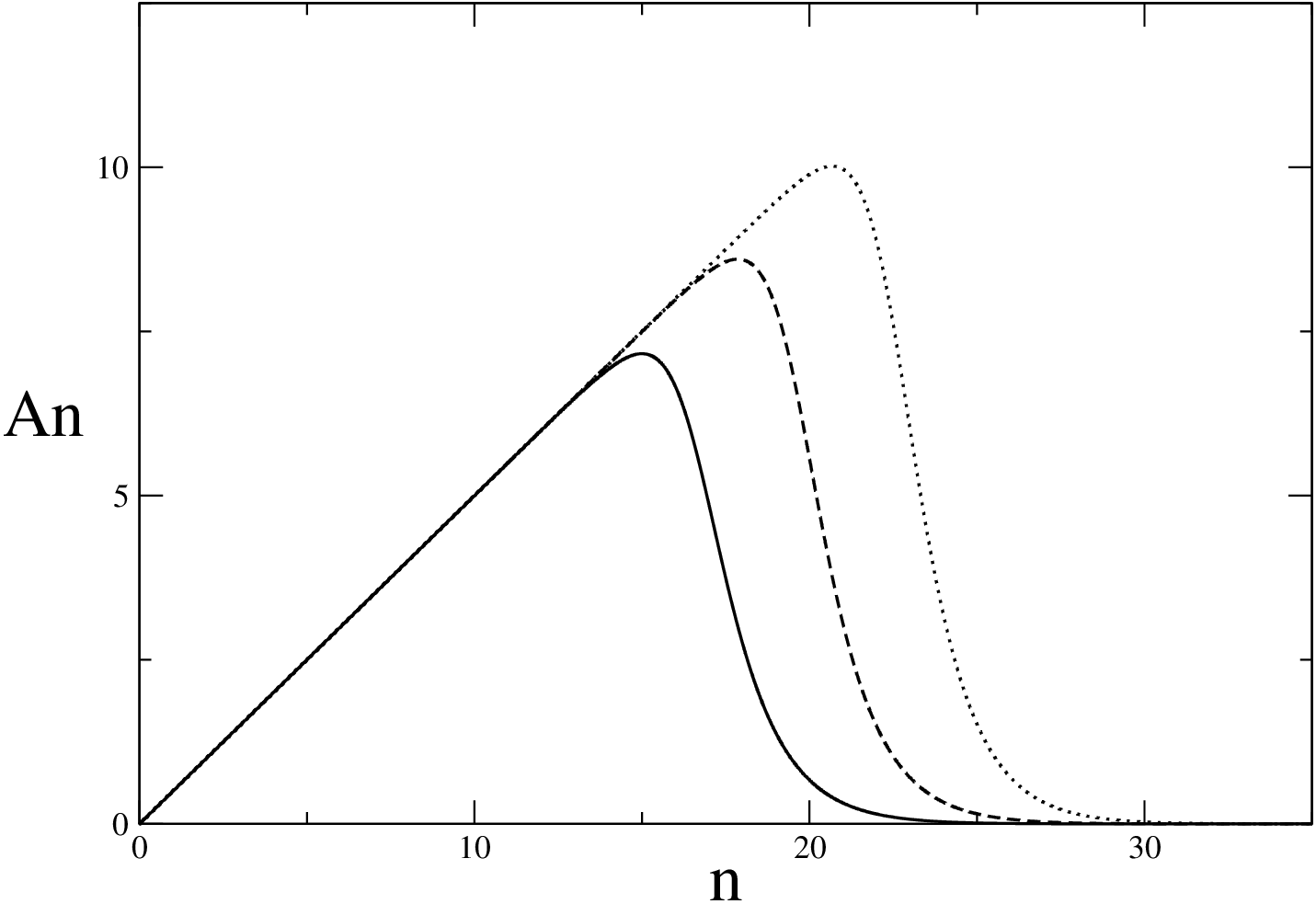}
\caption{Nonmonotonic behaviour of $A_n$ as a function of $n$ for three different values of the forcing frequency:
$\omega =10^{-3}$ (solid line), $\omega =10^{-4}$ (dashed line), and $\omega =10^{-5}$ (dotted line). We can see how the resonance appears for higher values of $n$ and becomes stronger as the forcing frequency decreases.}
\label{resonance}
\end{figure}

Now we can treat our problem as a two-state system governed by the master equation
$\dot{P}_{\pm}(t)=-W_{\mp}(t)P_{\pm}(t)+W_{\pm}(t)P_{\mp}(t)$,
where $P_+$ ($P_-$) denotes the probability that the system occupies the state $q=L^*$ (q=0) at time $t$, and $W_{+,-}$ denote the transition probability densities, given by
\begin{subequations}
\begin{eqnarray}
\label{trans1}
W_-=W(L^* \to 1)=\frac{1}{\tau(1- n A_2 \cos(\omega n^{-2} t))}\approx r (1+ n A_2 \cos(\omega n^{-2} t)), \\
W_+=W(0 \to 1)=\frac{1}{\tau(1+n A_1 \cos(\omega n^{-2} t))}\approx r (1- n A_1 \cos(\omega n^{-2} t)),
\label{trans2}
\end{eqnarray}
\end{subequations}
where $r=\tau^{-1}$. From now on, our derivation of the stochastic resonant behaviour will be parallel to that of
Ref.~\cite{jung}, so we will not do the calculations in detail and we refer the interested reader to this reference. We can use the normalization condition $P_++P_-=1$ together with the transition probability
densities Eqs.~(\ref{trans1}) and (\ref{trans2}) to solve the two-state master equation to first order in $A$. This solution can be used to obtain the system response $\left< q(t)|q(t_0),t_0 \right>=\int x \mathcal{P}(x,t|x_0,t_0)dx$, where $\mathcal{P}(x,t|x_0,t_0)=P_+(t) \delta(x-L^*)+ P_-(t) \delta(x)$. In the asymptotic limit we find
\begin{equation}
\label{response}
\lim_{t_0 \to -\infty} \left< q(t)|q(t_0),t_0 \right>=\frac{L^*}{2}-\frac{L^*}{2}(A_1+A_2)\frac{n^3r}{\sqrt{4r^2n^4+\omega^2}}
\cos \left[\frac{\omega}{n^2} t - \mathrm{arctan} \left( \frac{\omega}{2 r n^2} \right) \right],
\end{equation}
and we can appreciate that the amplitude of the periodic part of the system response undergoes a resonance for a finite value of $n$. To see this more clearly consider the $n$-dependent part of the amplitude of the system response
\begin{equation}
\label{an}
A_n=\frac{n^3r}{\sqrt{4r^2n^4+\omega^2}}=\frac{n^3(L^*-1)\exp(-nR)}{\sqrt{4(L^*-1)^2 \exp(-2nR)n^4+\omega^2}},
\end{equation}
where $R=S_0/n$ is a constant independent of $n$. This function attains its maximum at the value of $n$ solving the transcendental equation $4(L^*-1)^2n^4=\exp(2nR)(nR-3)\omega^2$. In order to get a more graphical idea of the phenomenon we have depicted $A_n$ versus $n$ in Fig.~(\ref{resonance}) for three values of the frequency
$\omega=10^{-3},10^{-4},10^{-5}$. In this figure one can see how the resonance becomes stronger and appears for higher values of $n$ as the forcing frequency decreases.

Let us now point out three final and very important remarks. The first is that the resonance is present only if the external signal is introduced modulating $K_4$, $K_3$, or $K_2$. A simple analysis reveals that modulating $K_1,K_2,K_3$ shifts the cubic dependence in $n$ in $A_n$ to $n^0,n^1,n^2$ respectively, so the possibility of resonance is lost (at least, if we do not introduce $n$-dependence in the external signal) for $K_1$. Also, this is not the unique example of stochastic resonance that might appear in a chemical system: different parameter values and reaction sets will provide new examples of this phenomenon. The assumptions made in this work were introduced to facilitate the analytical assessment of the problem. Finally, the improvement in the resonance observed in Fig.~(\ref{resonance}) for decreasing frequencies is not unbounded; if we assume too large values of $n$ then we could not consider the action~(\ref{spertur}) as a perturbation, and so the linear response theory would no longer be valid.

In summary, we have shown that it is possible to find stochastic resonance in reaction kinetics purely due to the presence of intrinsic noise generated by the discrete character of the reactants. We have presented our model as a population dynamics problem, a type of system that is usually subject to external periodic forces, as seasonal variation. A possible resonant coupling between phenotype selection in a biological species and periodic environmental evolution has been suggested recently~\cite{dunkel}, so it would be very interesting to know if not only the phenotype but also the population itself can undergo some sort of stochastic resonance, and in particular the one reported here. This could have a serious impact on the possibility of extinction of a population, since, as we have seen, internal fluctuations can drive a system from a state with a large population to an empty state.

This work has been partially supported by the Ministerio de Educaci\'{o}n y Ciencia (Spain) through Projects No. EX2005-0976 and FIS2005-01729.

\end{document}